\documentclass[12pt]{iopart}

\usepackage[utf8]{inputenc}

\expandafter\let\csname equation*\endcsname=\relax
\expandafter\let\csname endequation*\endcsname=\relax

\usepackage{amsmath}
\usepackage{amsthm}
\usepackage{graphicx}
\usepackage{multirow}
\graphicspath{{figs/}}
\usepackage{amssymb}
\usepackage{lmodern}
\usepackage{microtype}

\begin{document}

\newcommand \be {\begin{equation}}
\newcommand \ee {\end{equation}}
\newcommand \bea {\begin{eqnarray}}
\newcommand \eea {\end{eqnarray}}
\newcommand \ve {\varepsilon}

\newcommand{\iid}{\mathrm{iid}}

\newcommand{\R}{\mathbb{R}}
\newcommand{\N}{\mathbb{N}}
\newcommand{\D}[1]{\overset{\mathcal{D}} {#1}}
\newcommand{\conv}[2] {\underset{#1 \rightarrow #2} {\rightarrow}}

\newcommand{\action}[2]{\vphantom{#2}^{#1}#2}
\newcommand{\T}{\mathcal{T}}
\newcommand{\F}{\mathcal{F}}
\newcommand{\Eqref}[1]{Eq.~\eqref{#1}}

\newcommand{\Condref}[1]{Eq.~\eqref{#1}}

\newcommand{\U}{\mathcal{U}}

\newcommand{\Q}{Q}

\newcommand{\Inv}[1] {#1^{-1}}

\newcommand{\CL} {\mathcal {CL}}
\newcommand{\NSL} {\mathcal {PL}}
\newcommand{\simD}[2]{\underset{#1 \rightarrow #2} {\sim}}
\newcommand{\simInf}[1]{\simD{#1}{+\infty}}

\newcommand{\lgamma}{\Gamma}
\newcommand{\llambda}{\Lambda}
\newcommand{\ldelta}{\Delta}
\newcommand{\lQ}{\mathcal{Q}}
\newcommand{\NRV}[1]{\mathcal{NRV}_{#1}}
\newcommand{\CF}{\overline{F}}
\newcommand{\pder}[1]{\, \partial_{#1} }

\title[Renormalization for extreme value statistics]{Renormalization flow for extreme value statistics of random variables raised to a varying power}

\author{Florian Angeletti, Eric Bertin and Patrice Abry}

\address{Universit\'e de Lyon, Laboratoire de Physique,
Ecole Normale Sup\'erieure de Lyon, CNRS,
46 all\'ee d'Italie, F-69007 Lyon, France}

\begin{abstract}
Using a renormalization approach,
we study the asymptotic limit distribution of the maximum value
in a set of independent and identically distributed random variables
raised to a power $q_n$ that varies monotonically with the sample size $n$. Under these conditions, a non-standard class of max-stable limit distributions, which mirror the classical
ones, emerges. Furthermore a transition mechanism between the classical and the non-standard limit distributions is brought to light. If $q_n$ grows slower than a characteristic function $q^*_n$, the standard limit distributions are recovered, while if $q_n$ behaves asymptotically as $\lambda q^*_n$, non-standard limit distributions emerge. 
\end{abstract}

\pacs{05.10.Cc, 02.50.-r, 05.40.-a}
%keywords :
%extreme value statistics, renormalization, power transformations, non-standard limit distribution, attraction domain. 

\section{Introduction}

Extreme value statistics, that is the statistics of the largest value
in a set of random variables, has attracted a lot of attention in many
different fields, from probability theory \cite{Galambos:1978}
to physics --where fecund interactions with disordered systems
\cite{BouchaudMezard:1997,CarpentierLeDoussal:2001,FyodorovBouchaud:2008,FyodorovLeDoussal:2009}, as well as random walks and interface fluctuations
\cite{GyorgyiETAL:2003,LeDoussalMonthus:2003,MajumdarComtet:2004}
have recently flourished-- hydrology \cite{KatzParlangeNaveau:2002},
finance \cite{EmbrechtETAL:1997,Longin:2000} or
engineering \cite{Gumbel:1958}.
For independent and identically distributed (iid) random variables,
asymptotic distributions have been known for long
\cite{FisherTippett:1928,Gnedenko:1943,Gumbel:1958,Galambos:1978}.
Depending on the tail of the parent distribution (from which the variables
in the set are drawn at random), three different distributions emerge.
For parent distributions with a tail decaying faster than any power law,
the limit distribution is the well-known Gumbel one (which also found
interesting applications outside the field of extreme values
\cite{CluselBertin:2008}).
If the parent distribution decays as a power law close to infinity,
the so-called Fr\'echet distribution is obtained, while if it decays
algebraically close to an upper bound, the Weibull distribution is reached.

In spirit, these results bear some similarities with the Central Limit
Theorem, which addresses a similar issue for the problem of random sums
instead of extreme values. Interestingly, it has been shown recently
that the limit distribution of sums can be modified by raising the
summed variables to a power that diverges with the number of terms
in the sum \cite{BenArous:2005}. Such a problem is actually motivated
by the physics of disordered systems, as it can be interpreted
as the partition function of the Random Energy Model \cite{Derrida:1981},
one of the simplest disordered models
--which led to recent developments in relation to extreme value
statistics \cite{CarpentierLeDoussal:2001,FyodorovBouchaud:2008,FyodorovLeDoussal:2009}.
This problem also exhibits interesting connections to empirical moment
estimation in signal processing and multifractal analysis \cite{Angeletti:2011}.
It is then natural to wonder whether such a procedure, namely raising
the random variable to a power increasing with the sample size,
could generate some non-standard distributions as far as extreme values are concerned.
In terms of Random Energy Model, this would mean considering the statistics
of the maximum value of the Boltzmann weights (which add up to the partition
function).
A related, but perhaps more concrete, physical example is the statistics
of the largest trapping time in a trap model \cite{Bouchaud:1992},
in which particles are trapped in deep energy wells and can escape
only through thermal activation.
These extreme times are known to play an important role in this context.
In the limits of low temperature and large number of traps, the statistics
of the largest trapping time could depend on the way the two limits are taken.

This issue has been recently addressed in the mathematical literature
\cite{Bogachev:2007}, following the work by Ben Arous and coworkers on the
problem of sums, obtaining precise results about a transition between the Gumbel attraction domain and the Fr\'echet attraction domain for a specific class of distributions.
In addition, this problem has some connections with the question of the existence
of different limit distributions using power rescaling procedures
\cite{Pantcheva:1985,Mohan:1993,Grigelionis:2004,Ravi:2010,Calvo:2011}.

In the present contribution, we address the general issue of the limit distribution
of the maximum value in a set of random variables raised to a power exponent
diverging with sample size, by generalizing the renormalization group approach
recently introduced to deal with finite size effects in standard
extreme value statistics \cite{GyorgyiETAL:2008,GyorgyiETAL:2010,Bertin:2010}.
For exponents increasing as a power law of the sample size, we find
non-standard limit distributions, which turn out to be related
by an exponential change of variables to the standard limit
distributions. We clarify this surprising relationship using a simple
argument based on the behaviour of the rescaling factors.

\section{Problem statement}

Starting from a set of iid variables $(W_i)_{i=1,\ldots,n}$, we consider
the maximum $M_n^W$ in the set, namely:
\be
M_n^W =\max \{ W_1,\dots W_n\}.
\ee
Classical extreme value theorems yield asymptotic convergence results for $M_n^W$
as a function of the behaviour of the tail of the probability distribution of $W$.
More precisely, there exist two sequences $\alpha_n$ and $\beta_n$ such that
the cumulative distribution of the rescaled random variable $Y_n=(M_n^W-\beta_n)/\alpha_n$
converges to the limit cumulative distribution $\mathcal{F}_{\zeta}(y)$ defined by

\begin{equation}
\label{eq:L_zeta}
\mathcal{F}_{\zeta}(y)=
\begin{cases}
e^{-(1+\zeta y)^{-1/\zeta}} & \quad \mathrm{if} \; \zeta>0 \quad (y>-1/\zeta)\\
e^{- e^{-y}} & \quad \mathrm{if} \; \zeta=0\\
e^{-(1+\zeta y)^{-1/\zeta}} & \quad \mathrm{if} \; \zeta<0 \quad (y<-1/\zeta)\\
\end{cases} 
\end{equation}
The case $\zeta>0$ corresponds to variables $W_i$ with a distribution decaying
as a power law at infinity, while the case $\zeta<0$ rather corresponds
to a power-law decay close to an upper bound. Finally, the value $\zeta=0$
is obtained for distributions decaying faster than any power law (either
at infinity or close to an upper bound)
\cite{FisherTippett:1928,Gnedenko:1943,Gumbel:1958,Galambos:1978}.

We wish to investigate whether a $n$-dependent transformation
of the variables $W_i$ may lead to asymptotic distributions different from
the present ones. We are especially interested in power transformations of the form $U_{i,n}=W_n^{q_n}$ where $q_n$ depends on $n$, but begin by considering the general class of transformations 
\begin{equation}
U_{i,n}=\omega_n^{-1}(W_i), \qquad i=1,\ldots n,
\end{equation}
where $\omega_n$ consists in an increasing bijective function \footnote{Considering decreasing bijective functions is another possibility which would require only minor changes.}. 
One can express the cumulative distribution of $U$ as :
\begin{equation}
F_{U,n}(u) = F_W \big(\omega_n(u)\big).
\end{equation}
In the power transformation case, this definition of $\omega_n(u)$ leads to 
\be
\omega_n(u)= u^{1/q_n}, \qquad u>0.
\ee
This transformation is reminiscent of the study in \cite{BenArous:2005} concerning the behaviour of sums of random exponentials and notably the failure of the classical central limit theorem for rapidly growing powers. 

On the one hand, if $\omega_n$ varies sufficiently slowly as a function of $n$, it is expected that the transformation $\omega_n$ does not affect the limit distribution of the maximum.
On the other hand, for well-chosen transformations, it should be possible to attain new types of limit distributions.
Let us consider the transformed maximum $M_n^U = \max \{ U_{1,n},\dots U_{n,n}\}.$
The cumulative distribution $F_n(m)$ of $M_n^U$ can be expressed in terms of the
cumulative function $F_W$ of the variable $W$ as : 

\begin{equation}
\label{eq:F_omega}
F_n(m) = F_W\big(\omega_n(m)\big)^n . 
\end{equation}
In the following section, we devise a renormalization group formulation
of Eq.~(\ref{eq:F_omega}), which allows us to derive in a straightforward way
the possible fixed point distributions.

\section{Renormalization approach \label{sec:Renormalisation}}

\subsection{Renormalization transformation and standardization conditions}

Following Refs.~\cite{GyorgyiETAL:2010,Bertin:2010}, we introduce the functions
$g_n(m)= -\ln [-\ln F_n(m)]$ as well as $g_W(w)= -\ln [-\ln F_W(w)]$,
and recast \Eqref{eq:F_omega} into the form
\begin{equation}
\label{eq:g_omega}
g_n(m) = g_W\big(\omega_n(m)\big) - \ln n.
\end{equation}
As in the case of standard convergence theorems,
it is useful in order to converge to a non-degenerate limit distribution
to rescale the maximum value $M_n^U$ through $X_n=(M_n^U-b_n)/a_n$,
where $a_n$ and $b_n$ are chosen so as to meet some specific conditions
(for instance, fixing the values of the first two moments).
In addition, it is also convenient to consider $n$ as a real variable rather than
an integer one and to define the variable $s=\ln n$.
We thus assume that $\omega_n(m)$ can be extended to real values of $n$,
and we define the function $\omega(m,s)=\omega_{e^s}(m)$.
Altogether, one obtains from \Eqref{eq:g_omega} the following evolution equation,
in terms of the variable $X \equiv X_{e^s}$:
\be
g(x,s) = g_W\big(\omega(a(s)x+b(s),s)\big) - s,
\ee
where $\exp[-\exp(-g(x,s)]$ is the cumulative distribution of $X$.
In order to determine $a(s)$ and $b(s)$, one needs to impose
'standardization' conditions on $g(x,s)$. Such constraints
are arbitrary to some extent, and may differ depending whether one is interested
in practical problems or in theoretical approaches. In practical applications,
fixing some moments of the distribution (e.g., the first two moments) may be
convenient. In contrast, it turns out that for theoretical purposes,
fixing the value of $g(x,s)$ and of its derivative $\partial_x g(x,s)$
at a given value of $x$ is an easier condition to implement
\footnote{Throughout the paper, we use the notations $\partial_x \equiv \partial/\partial x$
and  $\partial_s \equiv \partial/\partial s$.}.
We thus choose the same conditions as in \cite{Bertin:2010}, namely
\begin{equation}
\label{eq:normalization:condition}
\begin{gathered}
g(0,s)= 0, \\
\partial_x g(0,s)= 1.
\end{gathered}
\end{equation}
These conditions imply
\begin{eqnarray}
\label{eq:normalization:res1}
&& g_W\big(\omega(b(s),s)\big) = s, \\
&& a(s)= b'(s) + \frac{\partial_s \omega(b(s),s)}{\partial_m \omega(b(s),s)},
\label{eq:normalization:res2}
\end{eqnarray}
with $b'(s)$ the derivative of $b(s)$ and $\partial_m \omega$ indicates the derivative with respect to the first argument of $\omega(m,s)$.
In order to simplify the expression of $g_W$, we contract the transformation
$\omega(x,s)$ and the rescaling operation into a single transformation $\T(x,s)$:

\begin{equation}
\label{eq:def:T}
\T(x,s)= \omega(a(s)x+b(s),s),
\end{equation}
which leads to
\begin{equation}
\label{eq:RG-T}
g(x,s)= g_W(\T(x,s) ) - s.
\end{equation}

\subsection{Partial differential equation for the flow}

The functional equation (\ref{eq:RG-T}) can be converted into a partial
differential equation.
We first differentiate $g(x,s)$ with respect to $x$ and $s$:
\begin{eqnarray}
\label{eq:PDE:dx}
\partial_x g(x,s) &=& g_W'(\T(x,s))\, \partial_x \T(x,s),\\
\label{eq:PDE:ds}
\partial_s g(x,s) &=& g_W'(\T(x,s))\, \partial_s \T(x,s) -1.
\end{eqnarray}
Reinjecting \Eqref{eq:PDE:dx} into \Eqref{eq:PDE:ds} to eliminate $g_W'$, we obtain
\begin{equation}
\label{eq:PDE:total}
\partial_s g(x,s)= \U(x,s)\, \partial_x g(x,s) -1,
\end{equation}
where we have defined
\be \label{eq:def:U}
\U(x,s) =  \frac{\partial_s \T(x,s)}{\partial_x \T(x,s)}.
\ee
The function $\U(x,s)$ can be expressed explicitly using Eqs.~(\ref{eq:def:T})
and (\ref{eq:def:U}).
We start by computing $\partial_x \T$ and $\partial_s \T$
(in order to lighten the notations, we drop in the following the explicit $s$
dependence of the parameters $a(s)$ and $b(s)$):
\begin{align}
\partial_x \T(x,s) &= a\, \partial_m \omega(a x + b,s),\\
\partial_s \T(x,s) &= (a' x + b')\, \partial_m \omega(a x+b,s) +
\partial_s \omega(a x+b,s).
\end{align}
We finally obtain
\begin{equation}
\label{eq:PDE:quotient}
\U(x,s) = \frac{a'}{a}\, x + \frac{b'}{a} + \frac{1}{a}\frac{\partial_s \omega(ax+b,s)}{\partial_m \omega(ax+b,s)}.
\end{equation}
From now on, we focus on the case of a power-law transformation
$\omega(m,s) = m^{1/q(s)}$, with $m>0$. One has
\begin{eqnarray}
\partial_m \omega(m,s) &=& \frac{\omega(m,s)}{q(s) m},\\
\partial_s \omega(m,s) &=& -\frac{q'(s) \ln m}{q(s)^2} \,\omega(m,s),
\end{eqnarray}
and $\U(x,s)$ reads
\begin{equation}
\label{eq:PDE:quotient:q}
\U(x,s) = \frac{a'}{a} x + \frac{b'}{a} - \frac{q'}{q} \left(x + \frac{b}{a}\right)
\ln(ax+b).
\end{equation}

Taking into account the standardization condition (\ref{eq:normalization:res2}),
which reads, in the case of a power-law transformation $\omega(m,s)$,
\be \label{std2-power-law}
a = b'-\frac{q'}{q} b\ln b,
\ee
one can rewrite Eq.~(\ref{eq:PDE:quotient:q}) as
\be
\label{eq:PDE:quotient:q2}
\U(x,s) = \left( \frac{a}{b} +\partial_s \ln \frac{a}{b} \right) x
+1 -\frac{q'}{q}\, \left(x+\frac{b}{a} \right)
\; \ln\left( \frac{a}{b}x+1\right).
\ee
If one defines
\begin{equation}
\label{eq:PDE:params}
 \lambda(s)= \frac{a(s)}{b(s)}, \quad \delta(s)=\partial_s \ln \lambda(s), \quad \gamma(s)=\lambda(s)+\delta(s), \quad \Q(s) = \frac{q'(s)}{q(s)},
 \end{equation}
it is possible to rewrite \Eqref{eq:PDE:quotient:q2} in a more compact way as
\begin{equation}
\label{eq:PDE:named}
\U(x,s)= 1+ \gamma(s)  x -\Q(s) \left(\lambda(s) x+ 1 \right)\; \frac{ \ln ( \lambda(s) x+1)} {\lambda(s)} .
\end{equation}

\subsection{Fixed point distributions}

A stationary solution, that is a solution of Eq.~(\ref{eq:PDE:total})
satisfying $\partial_s g(x,s)=0$, can
be obtained on condition that $\U(x,s)$ is independent of $s$, namely
$\U(x,s)=\U(x)$.
In this case, the stationary solution $g(x)$ is determined by integrating
the differential equation
\be \label{eq:PDE:stationary}
g'(x) = \frac{1}{\U(x)},
\ee
with the condition $g(0)=0$.

We now investigate under which condition $\U(x,s)$ becomes independent of $s$.
\subsection{Case $\Q=0$: recovering standard limit distributions}

In the case $\Q=0$, one has
\begin{equation}
\label{eq:PDE:quotient:dyn:Q0}
\U(x,s) = 1 + \gamma(s) x.
\end{equation}
The condition $\partial_{s} \U=0$ yields that $\gamma(s)$ must be equal
to a constant $\gamma$.
Consequently,
\begin{equation}
g'(x)=\frac{1}{1+\gamma x}.
\end{equation}
Taking into account the standardization condition (\ref{eq:normalization:condition}),
we obtain the fixed point function
\begin{equation}
\label{eq:lim:0}
g(x)=\frac{1}{\gamma} \ln(1+\gamma x).
\end{equation}
Reformulating this result in terms of the cumulative distribution
$ F(x)=\exp[-\exp(-g(x))]$, one recovers the standard limit distributions
given in \Eqref{eq:L_zeta},
\begin{equation}
\label{eq:lim:F:0}
F(x) = \mathcal{F}_{\gamma}(x)= \exp\left[-(1+\gamma x)^ {-\frac{1}{\gamma} } \right], \qquad
1+\gamma x >0,
\end{equation} 
where $\gamma$ plays the role of the parameter $\zeta$.
Hence classical limit distributions are retrieved in the case where the power
$q(s)$ is a constant, namely $q(s)=q_0$. This result was expected: if $X$ belongs to the attraction domain of $\mathcal{F}_\zeta$, $X^{q_0}$ either belongs to the attraction domain of
$\mathcal{F}_{\zeta/q_0}$ for $\zeta >0$, or to the attraction domain of $\mathcal{F}_\zeta$
otherwise.

\subsection{Case $\Q\neq 0$: emergence of non-standard stable distributions}

We now turn to the case $\Q \ne 0$. From Eq.~(\ref{eq:PDE:named}), it is clear that $\U(x,s)$ is independent of $s$ if $\Q(s)$ and $\lambda(s)$ are constants,
$\Q(s)=\Q$ and $\lambda(s)=\lambda>0$.
Hence stationary solutions only exist if $q(s)$ is of the form $q(s)= K\, e^{\Q s}$,
with $K>0$ a real constant.
Inserting $\Q$ and $\lambda$ in \Eqref{eq:PDE:quotient:q2}, $\U(x)$ takes the form
\be
\label{eq:PDE:U:stationary}
\U(x) = 1+\lambda x -\frac{\Q}{\lambda} (1+\lambda x) \ln (1+\lambda x).
\ee
Combining \Eqref{eq:PDE:U:stationary} with Eq.~(\ref{eq:PDE:stationary}) leads to
\be
\label{Eq:gprime-Q-gamma}
g'(x) = \frac{1}{(1+\lambda x) \left(1-\frac{\Q}{\lambda}\ln (1+\lambda x) \right)}.
\ee
By definition, $g(x)$ has to be an increasing function of $x$ so that $g'(x)\ge 0$,
which implies that $x$ belongs to a restricted range of values,
$x_{\min} < x < x_{\max}$.
The bounds $x_{\min}$ and $ x_{\max}$ are determined by the conditions
$1+\gamma x>0$ and $1-\frac{\Q}{\lambda}\ln (1+\lambda x)>0$.
Assuming $\Q>0$, one finds

\begin{equation}
x_{\min} = -\frac{1}{\lambda} \, , \quad \;
x_{\max} = \frac{1}{\lambda} \left( e^{\lambda/\Q}-1 \right) \, , \qquad \lambda >0.
\end{equation}
Similarly, for $\Q<0$,
\begin{equation}
x_{\min} = \frac{1}{\lambda} \left( e^{\lambda/\Q}-1 \right)\, , \quad \;
x_{\max} = +\infty\, ,
 \qquad \lambda >0.
\end{equation}

\noindent Coming back to Eq.~(\ref{Eq:gprime-Q-gamma}), this equation can be integrated into
\be
\label{eq:lim:Q}
g(x) = -\frac{1}{\Q} \, \ln \left(1-\frac{\Q}{\gamma}\ln (1+\gamma x) \right), \qquad x_{\mathrm{min}} < x < x_{\mathrm{max}},
\ee
also taking into account the condition $g(0)=0$.
The corresponding cumulative distribution $\F_{\lambda,\Q}(x)$ reads
\be
\label{eq:lim:F:Q}
\F_{\lambda,\Q}(x) = \exp\left[- \left(1-\frac{\Q}{\lambda}\ln (1+\lambda x)
\right)^{1/\Q} \right], \qquad x_{\min} < x < x_{\max},
\ee
which generalizes the standard extreme value distributions;
for $Q=1$, this expression reduces to a power law on the interval
$[x_{\min}, x_{\max}]$.
Note that the expression (\ref{eq:lim:F:Q}) of $\F_{\lambda,\Q}(x)$ converges,
in the limit $\Q \rightarrow 0$, to the standard distribution $\F_{\lambda}(x)$
given in Eq.~(\ref{eq:lim:F:0}).
In addition, it is interesting to note that in the limit $\lambda \rightarrow 0$,
the cumulative $\F_{\lambda,\Q}$ converges to $\F_{-\Q}$.  

\begin{figure}
\center{
\includegraphics{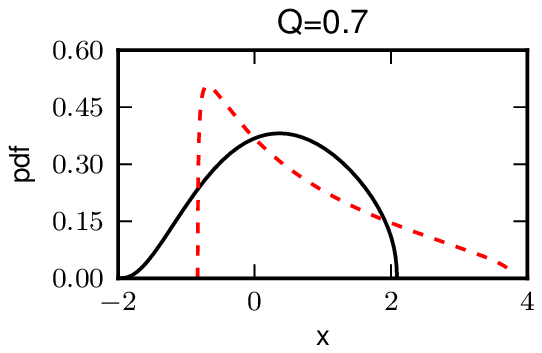} \includegraphics{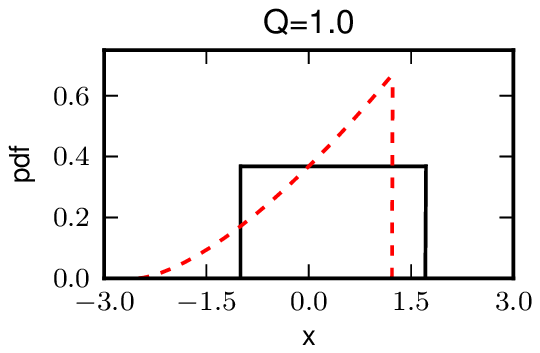}
\includegraphics{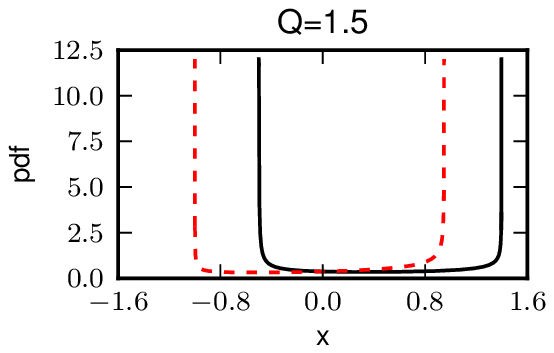} \includegraphics{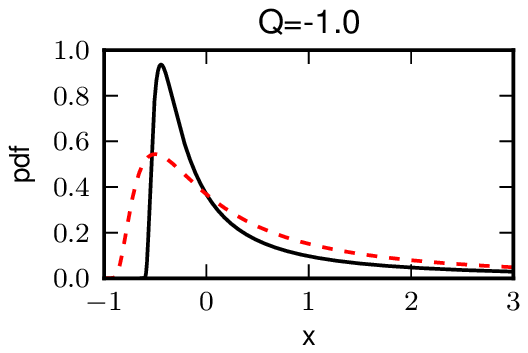}

\caption{(Colour online) Probability density function $p_{\lambda,\Q}(x)$ associated to
the generalized extreme distribution $\F_{\lambda,\Q}(x)$.
Top left: $\Q=0.7$, solid black line $\lambda=0.5$, dashed red line $\lambda=1.2$. 
Top right: $\Q=1$, solid black line $\lambda=1$, dashed red line $\lambda=0.4$.
Bottom left: $\Q=1.5$, solid black line $\lambda=2$, dashed red line $\lambda=1$.
Bottom right: $\Q=-1$, solid black line $\lambda=1$, dashed red line $\lambda=0.01$.
}
\label{fig:pdf}
}
\end{figure}

Remarkably, these generalized extreme value distributions are closely related to the standard ones. In term of cumulative, expression (\ref{eq:lim:F:Q}) implies that 
\begin{equation}
\label{eq:mirror}
\F_{\lambda,\Q}(x) = \F_{-\Q}\big( \ln (1+\lambda x) / \lambda \big).
\end{equation}
Incidentally, setting $\zeta=-\Q$, Eq.~(\ref{eq:mirror}) implies that if
$Z$ is a random variable with cumulative $\F_\zeta$, then the cumulative of
$[\exp{(\lambda Z)}-1]/\lambda$ is $\F_{\lambda,-\zeta}$.
Note that Eq.~(\ref{eq:mirror}) is also valid for $\Q=0$,
and corresponds in this case to a known relation between the Gumbel
and the Fr\'echet distributions. 

It should also be noted that for $\Q>0$ these limits distributions are
equivalent, up to an affine transformation, to the non-standard-distribution
obtained in \cite{Calvo:2011} using power transformations as rescaling factors.
This suggests that $q(s)$ plays a role akin to a rescaling factor.
However, let us emphasize that our approach is different in spirit
from that of Ref.~\cite{Calvo:2011}. In the latter, the power $q_n$
is used as an adjustable rescaling parameter allowing the transformed
distribution to converge to a non-degenerate limit.
In the present paper, we consider the power $q_n$ as a given function,
and we study the non-degenerate limit distributions that are obtained
through a standard affine rescaling of the data.
It is then a priori not obvious that the same stable distributions should
emerge from the two procedures.

Deriving the probability density function $p_{\lambda,\Q}(x)$ of these generalized extreme value distributions yields the expression
\begin{equation}
\label{eq:Fl:pdf}
p_{\lambda,\Q}(x) =\frac{1}{1+\lambda x} \left[ 1-\frac{\Q}{\lambda} \ln(1+\lambda x) \right]^{\frac{1}{\Q}-1} \exp\left[- \left(1-\frac{\Q}{\lambda}\ln (1+\lambda x)
\right)^{1/\Q} \right].
\end{equation}
As depicted by Fig.~(\ref{fig:pdf}), this family of probability density functions
has a non-trivial behaviour near its bounds. For $Q>0$, the support of $p_{\lambda,\Q}$ is
$(-1/\lambda,[\exp(\lambda/Q)-1]/\lambda)$. If we consider the asymptotic behaviour
in the limit $x\equiv -1/\lambda+\epsilon$, with $\epsilon \rightarrow 0$, we obtain
\begin{equation}
\label{eq:Fl:pdf:asympInf}
p_{\lambda,\Q}\left(-\frac{1}{\lambda}+\epsilon\right) \approx \frac{1}{\epsilon} \left[\frac{\Q}{\lambda} |\ln(\epsilon)| \right]^{\frac{1}{\Q}-1} e^{ |\frac{\Q}{\lambda} \ln \epsilon| ^ {1/\Q} }.
\end{equation}
Therefore, we have a crossover at $\Q=1$ where the probability density corresponds to a power law of exponent $1/\lambda-1$. For $Q>1$, $p_{\lambda,\Q}(x)$ diverges
when $x \rightarrow -1/\lambda$ faster than any power functions, whereas for $Q<1$
and all $n$, $p_{\lambda,\Q}(x)$ and all its derivatives converge to $0$.

Similarly, in the neighborhood of the upper bound $x\equiv [\exp(\lambda/Q)-1]/\lambda -\epsilon$, we have 
\begin{equation}
\label{eq:Fl:pdf:asympSup}
p_{\lambda,\Q}\left(x_{\max} -\epsilon \right) \approx 
e^{-\lambda/Q} \left(\Q e^{-\lambda/Q} \epsilon \right) ^{\frac{1}{\Q}-1},
\end{equation}
and a singularity appears at the upper bound of the distribution for $Q>1$.

Conversely, for $\Q<0$, $(\exp(\lambda/Q)-1)/\lambda$ becomes the lower bound
of this distribution, and using $ x\equiv x_{\min} +\epsilon$ and $\epsilon \rightarrow 0 $ leads to
\begin{equation}
\label{eq:Fl:pdf:asympSup2}
p_{\lambda,\Q}\left(x_{\min} +\epsilon \right) \approx 
e^{-\lambda/Q} \left(|\Q| e^{-\lambda/Q} \epsilon \right) ^{\frac{1}{\Q}-1} \exp \left[  \left(|\Q| e^{-\lambda/Q} \epsilon \right) ^{\frac{1}{\Q}} \right].
\end{equation}
Since $\Q<0$, the exponential term is dominant and the probability density function and all its derivatives
converge to $0$ in the neighborhood of $x_{\min}$. On the other hand, for $x\rightarrow +\infty$, $F(x)$ behaves as the exponential of the power of a logarithm: 
\begin{equation} 
 F(x) \underset{x\rightarrow+\infty} {\approx} \exp\left[- \left|\frac{\Q}{\lambda} \ln x\right|^{-\frac{1}{\Q}} \right].
\end{equation}
There is therefore four distinct asymptotic behaviours for the function $p_{\lambda,Q}(x)$ depending on the value of Q ($Q<0$, $0<Q<1$, $Q=1$, $Q>1$). The function $p_{\lambda,Q}(x)$ is shown in Fig.~\ref{fig:pdf} for several values corresponding to these distinct domains.

\section{Attraction domain of the non-standard limit laws}
\label{sec:Forcing}

Having found non-standard asymptotic forms of the extreme value distribution in Section~\ref{sec:Renormalisation}, we now consider their domain of attraction and we use the partial differential equation of the flow to develop a heuristic description of these attraction domains.
This description provides a good approximation of the standard attraction domains
and gives us precise insights into the non-standard attraction domains.
Finally, these heuristic results are confirmed in section \ref{sec:transition}
using an independent approach which sheds some light on the relationship
between standard and non-standard laws.

Although the stationary equation resulting from (\ref{eq:PDE:stationary})
and (\ref{eq:PDE:U:stationary}) cannot be used
directly to describe the dynamic of the general equation (\ref{eq:PDE:total}),
it underlines at least the existence of two important control parameters
in the stationary case, namely $\Q$ and $\lambda$.

If we return to the general flow equation (\ref{eq:PDE:total}) and the general expression of $\U$ described by \Eqref{eq:PDE:named}, these two variables can be interpreted as an external forcing.
Heuristically, the convergence of \Eqref{eq:PDE:total} should be driven by the asymptotic behaviour of these parameters. Notably, if $\gamma(s)$ and $\Q(s)$ converge simultaneously toward the respective finite limits $\lgamma$ and $\lQ$, it is plausible to expect the transformed maximum to converge toward the limit law $\F_{\lgamma, \lQ}$. As a corollary, for laws belonging to a classical domain with parameter $\zeta$
in the case $\Q(s)=0$, one should obtain $\lgamma=\zeta$, which sets an interesting test case for our heuristic argument.

\subsection{Asymptotic behaviours of the forcing parameter}

 If the term $\Q(s)$ clearly denotes the external forcing due to the power transformation, the interpretation of $\lambda(s)$ and $\delta(s)$ is hazier. From its definition, 
\begin{equation}
\label{eq:forc:lambda}
\begin{aligned}
\lambda(s) &= \pder{s} \ln b(s) - q'(s) \ln b(s) \\
&=  \frac{q(s)} {g_W^{-1}(s) g_W' \left( g_W^{-1}(s) \right)}\\
&= q(s) \lambda_0(s).  
\end{aligned}
\end{equation}
It is thus possible to factorize $\lambda(s)$ into the factor $q(s)$ and a term $\lambda_0(s)$ depending only on the cumulative function of the parent distribution. Using the change of variable $s=g(x)$ and the relation $g(x)=-\ln(-\ln F(x) )$ leads to
\begin{equation}
\lambda_0(g(x)) = \frac{ |F(x) \ln F(x)| }{x F'(x)}. 
\end{equation}
One should note that $\lambda_0(s) > 0$ due to the properties of the cumulative function. Moreover, defining the complementary cumulative function $\CF(x)=1-F(x)$ leads to the asymptotic expression
\begin{equation}
\lambda_0(g(x)) \simD{x}{x_F} - \frac{\CF(x)}{x \CF '(x)}. 
\end{equation} 
The variable $\lambda_0(s)$ corresponds to the inverse of the local power exponent of the complementary cumulative function $\CF$ at point $x=g(s)$. If $\lambda_0(s)$ admits a non-zero finite limit $\llambda_0$ at $+\infty$, there exists a normalized slowly varying function $L$ \footnote{The slowly varying functions class gathers constants, logarithms and all functions satisfying the asymptotic relation: 
$ \forall \alpha>0,\quad L(\alpha x) \simInf{x} L(x)$.
A slowly varying function is said to be normalized if $\lim_{x\rightarrow +\infty } x L'(x) / L(x) =0$ \cite[p.~15]{Bingham:RegularVariation}.
Since we are only interested in the generic properties of
slowly varying functions, we most often use the same notation $L$ for any
function belonging to this class.} such that
\begin{equation}
\CF(x) = L(x) x^{-1/\llambda_0}.
\end{equation} 
In other words, the parameter $\llambda_0$ describes the power behaviour of the tail of the distribution. Since for $\Q(s)=0$, $\lambda(s)$ and $\lambda_0(s)$ are identical, laws belonging to the classical domains of attraction  provide enlightening examples of the asymptotic behaviour of $\lambda_0(s)$. Notably, if a law belongs to the Fr\'echet attraction domain of parameter $\zeta$, there is a slowly varying function $L$ such that
\begin{equation}
\label{eq:domain:Frechet}
\CF(x)=L(x) x^{-\frac{1}{\zeta}}.
\end{equation}
With the regularity assumption that $L$ is normalized, we obtain that $\llambda_0=\zeta$. 
Conversely, for a law belonging to the Weibull domain, we have a slowly varying function $L$ such that \begin{equation}
\label{eq:domain:Weibull}
\CF(x)=(x_F-x)^{-1/\zeta} L\big((x_F-x)^{-1} \big)
\end{equation}
where $x_F$ is the end point of the distribution. Under the assumption that $L$ is normalized, we have
\begin{equation}
\lambda_0(g(x)) \simD{x}{x_F} \frac{\zeta}{x_F}(x_F-x).
\end{equation}
The Gumbel domain of attraction is harder to characterize.
However, the ``exponential power'' laws 
\begin{equation} 
\label{eq:domain:ExpPow}
 \CF(x)=\exp(-x^{\alpha})
\end{equation} 
often constitute an interesting subset of this domain.
For such laws, a short calculation leads to
\begin{equation}
\lambda_0(g(x)) \simInf{x} \frac{x^{-\alpha}}{\alpha}.
\end{equation} 
In the general case, a law belongs to the Gumbel attraction domain, if and only if $1/\CF(x)$ is a rapidly varying function. Using an integral characterization of rapidly varying functions \cite[p.~178]{Bingham:RegularVariation}, it is possible to show, that under a regularity assumption, for any law belonging to the Gumbel attraction domain, $\lambda_0(s) \rightarrow 0$.  
Consequently, for laws in the Weibull and Gumbel attraction domains, we have $\llambda_0=0$. In summary, for laws belonging to a standard attraction domain
of parameter $\zeta$, one has $\llambda_0=\max(0,\zeta)$.

Similarly, $\delta(s)$ can be decomposed into
\begin{equation}
\label{eq:forc:delta}
\begin{gathered}
\delta(s)=\Q(s) + \delta_0(s),\quad
\delta_0(s) = \pder{s} \ln \lambda_0(s).
\end{gathered}
\end{equation} 
One interesting consequence of the previous equation is that the asymptotic behaviours of $\delta_0(s)$ and $\lambda_0(s)$ are entwined.
If we assume that both $\delta_0(s)$ and $\lambda_0(s)$ admit a finite limit when
$s$ goes to $+\infty$, with $\lim_{s \rightarrow \infty} \lambda_0(s)>0$,
then one necessarily has $\lim_{s \rightarrow \infty} \delta_0=0$. This situation corresponds
to the Fr\'echet class.
Conversely, if $\lim_{s \rightarrow \infty} \lambda_0(s)=0$ (a typical situation in the
Gumbel and Weibull classes), and if $\delta_0(s)$ admits
a finite limit $\ldelta_0$ when $s\rightarrow \infty$, then $\ldelta_0<0$. 
After some tedious calculations, one can expand the expression
(\ref{eq:forc:delta}) of $\delta_0(s)$ as
\begin{equation}
\begin{aligned}
\delta_0(g(x))&= |\ln F(x)| - \frac{F''(x) |\ln F(x)| }{F'(x)^2}-1-\lambda_0(g(x))\\
 &\simInf{s} \frac{\CF(x) \, \CF''(x)}{\CF'(x)^2}-1-\lambda_0(g(x)).
\end{aligned}
\end{equation}
Going on with our study of the classical domains, we consider a law belonging to the Weibull attraction domain with parameter $\zeta$ and slowly varying function $L$. Under the regularity assumption that both $L$ and $L'$ are normalized slowly varying function, it is possible to verify that
$\ldelta_0=\zeta$. The same computation for the ``exponential power'' laws
(a typical example of laws belonging to the Gumbel domain) leads to 
\begin{equation} \delta_0(g(x))\simInf{x}  -x^{-\alpha}, \end{equation}
and therefore, $\ldelta_0=0$.
In other words, the limits $\llambda_0$ and $\ldelta_0$ are respectively the positive and negative parts of $\lgamma_0=\lim_{s \rightarrow \infty} (\delta_0(s)+\lambda_0(s))$,
namely $\llambda_0=\max(0,\lgamma_0)$ and $\ldelta_0=\min(0,\lgamma_0)$.
For a law belonging to the standard attraction domain of parameter $\zeta$, we have as expected $\lgamma_0=\zeta$. So, for $\Q=0$, we have recovered the classical results concerning the attraction domain (ignoring the difference in regularity assumptions). And from this point, it is easy to extend this convergence result to the non-standard case ($\Q\ne0$). 
 
Piecing together \Eqref{eq:forc:lambda} and (\ref{eq:forc:delta}), $\gamma(s)$ can also be rewritten as
\begin{equation}
\label{eq:gamma-deltaQlambda}
\gamma(s)=\delta_0(s) + \Q(s) + q(s) \lambda_0(s). 
\end{equation} 
The previous expression outlines an interesting interplay between the tail behaviour of the parent distribution and the power transformation. The tail behaviour of the distribution is responsible for the parameter $\delta_0$ whereas $\Q(s)$ is directly derived from the choice of $q(s)$. The term $q(s) \lambda_0(s)$ represents the interaction between the two effects.

\subsection{Changing the attraction domain by varying $q(s)$}

If we consider a fixed parent law, the functions $\delta_0(s)$ and $\lambda_0(s)$ are then fixed. Therefore the only free parameter is $q(s)$. In this situation, the limit $\lgamma$ is determined by the asymptotic behaviour of $\lambda(s)=q(s) \lambda_0(s) $.
Let us introduce the limit $\llambda \equiv \lim_{s\rightarrow \infty} \lambda(s)$,
when it exists.
If we define a characteristic power scale by 
\begin{equation}
\label{eq:forcing:qstar}
q^*(s)= \frac{1}{\lambda_0(s)},
\end{equation}
we obtain
\begin{equation}
\llambda= \lim_{s\rightarrow \infty} \frac{q(s)}{q^*(s)}.
\end{equation}
In other words, taking the limit $s \rightarrow \infty$ in Eq.~(\ref{eq:gamma-deltaQlambda}),
one finds that $\lgamma$ is only dependent of $\ldelta_0$, $\lQ$ and of the limit ratio between $q(s)$ and the characteristic power scale $q^*(s)$. 
From this point, two different situations arise. On the one hand, a special case appears if $ \lim_s q(s)/q^*(s)=0$. Indeed, if $\lim_{s\rightarrow \infty} \lambda(s)=0$, the expression of $\U(x,s)$ can be linearised as  
\begin{equation}
\U(x,s)\approx 1+ \gamma(s) x -\Q(s) x .
\end{equation}
So if $\gamma(s)$ and $\Q(s)$ admit the finite limits $\lgamma$ and $\lQ$, we obtain a partial differential equation corresponding to a standard limit distribution with parameter $\zeta=\ldelta_0$. 
This implies that the Weibull and Gumbel domains are unaffected by such a transformation. However, for a law belonging to the Fr\'echet domain of parameter $\zeta$, we have $\llambda_0=\zeta$ and $\ldelta=0$. This means that $\llambda=0$ is only possible with $\lim_s q(s)=0$ and $\Q<0$. 
In other words, a decreasing power destabilizes all the Fr\'echet domains and any law belonging to a Fr\'echet domain will converge towards a Gumbel distribution once exposed to a vanishing power transformation.

On the other hand, non-standard limit laws appears when $q(s) \sim \llambda q^*(s)$. In this case, we have
\begin{equation}
\lgamma = \ldelta_0 + \lQ + \llambda.
\end{equation}  
Moreover, the relation $\Q(s)=q'(s)/q(s)$ leads to
\begin{equation}
\Q(s)= -\delta_0(s) + \frac{\lambda'(s)}{\lambda(s)}.
\end{equation}
So, if we assume that $\lim_{s\rightarrow +\infty} \lambda'(s)/\lambda(s)=0$, one obtains
that

\begin{equation}
\left\{
\begin{gathered}
\lQ= - \ldelta_0,\\
\lgamma= \lambda.
\end{gathered}
\right.
\end{equation}  

So using a power transformation with $q(s) \sim \lambda q^*(s)$ leads to the non-standard limit law $\F_{\lambda,-\ldelta_0}$. Moreover, it is possible to compute an analytic representation of $q^*_n$ by defining the error term $\epsilon(s) = Q(s) + \Delta_0$. Using \Eqref{eq:PDE:params} leads to the exact differential equation
\begin{equation}
\partial_s \ln q^*(s) = -\Delta_0 + \epsilon(s).
\end{equation}
Solving this equation results in
\begin{gather}
\label{def-Lstar}
q_n^* = L^*(n) n^{-\Delta_0},\\
\label{eq:qnstar:slow} L^*(n) = \exp \left( \int_{0}^{n} \frac{\epsilon(\ln t)}{t} dt \right).
\end{gather}
The factor $L^*(n)$ corresponds to a corrective term depending on the fine convergence structure of $\delta_0(s)$ and $\lambda(s)$. Moreover, since $\epsilon(\ln n) \rightarrow 0$, \Eqref{eq:qnstar:slow} corresponds exactly to the integral representation of a normalised slowly varying function. In other words, $L^*(n)$ is a slowly varying correction to the power behaviour of $q_n^*$, if $\ldelta_0 \ne 0$. However, if $\ldelta_0=0$, this slowly varying term becomes preponderant. Notably for Fréchet distributions, $\llambda_0=\zeta$ implies that  $q_n^* \approx 1/\zeta$. On the other hand, for the Gumbel attraction domain, $\lambda_0(s) \rightarrow 0$ implies that $L^*(n) \rightarrow +\infty$. For instance, in the case of the exponential power laws, we have $\delta_0(s)=\epsilon(s) \sim \frac{1}{s}$, therefore $q_n \approx K \ln n$ ($K>0$). Computing $\lambda_0(s)$ yields the more precise expression
\begin{equation}
q^*_n \approx \alpha \ln n.
\end{equation}

Consequently, using the right power transformation ($q_n = \llambda L^*(n) n^{|\zeta|}$), a law belonging to the Weibull attraction domain can be forced to converge into a non-standard limit distribution $\F_{\lambda,-\ldelta_0}$. On the other hand, after applying a slowly varying power transformation ($q_n = \llambda \alpha \ln n$ for the exponential power laws) to a law belonging to the Gumbel attraction domain, one can force the convergence towards a Fr\'echet limit law. This is perfectly consistent with the (more precise) result obtained in \cite{Bogachev:2007} concerning the transition between the Gumbel domain and the Fr\'echet domain for a specific power transformation. 

However, considering parent distributions from standard domains only leads to non-standard distributions with positive $\lQ$. The domains of attraction of non-standard laws with negative $\lQ$ correspond to parent laws which do not belong to the standard attraction domains but can be ``renormalised'' using the decreasing power law.
One example of such laws would be the
logarithm-power law with 
\begin{equation}
\CF(x)=1-L(\ln x)(\ln x)^{-\frac{1}{\alpha}},
\end{equation}
where both $L$ and its derivative $L'$ are normalized slowly varying functions and $\alpha>0$.
A short calculation shows that $\lambda_0(g(x)) \sim \alpha \ln x$ and $\ldelta_0= \alpha$.
Thus these laws belong to the attraction domain of $\F_{\lambda,-\alpha}$ with $\alpha>0$ and $\lambda>0$. The associated power scale correspond to a decreasing power functions $q^*_n=L^*(n) n^{-\alpha}$.  Going further, with a power transformation decreasing faster than $n^{-\alpha}$, these logarithm-power law converge towards a Fr\'echet distribution of parameter $\alpha$.

\begin{table}
\newcommand{\tmr}[1]{\multirow{2}{*}{#1}}
\newcommand{\tmc}[3]{\multicolumn{#1}{#2}{#3}}
\newcommand{\mcmr}[3]{\multicolumn{#1}{#2}{\tmr{#3}}}
\center{
\begin{tabular}{|c|c|c|c|c|c|c|c|c|}
\hline
Parent law & \tmc{8}{|c|}{Functional dependence of $q_n$ }\\
\cline{2-9} 
     & \tmc{3}{|c|}{~~~~~~~~$L(n) n^{\Q},\quad\Q<0 $~~~~~~~~} & $\lambda$ & $L_d(n)$ & \tmc{3}{|c|}{~~~~~~~~$ L(n) n^{\Q},\quad\Q>0 $~~~~~~~~}    \\ 
\hline

Weibull dom. &                     \mcmr{5}{c}{\hspace{1cm}$\F_\zeta$}                &~~~   &  $Q=|\zeta|,\, L=\lambda L^*$             &\tmr{None}\\
 $\zeta<0$, $L^*(n)$     &                     \tmc{5}{c}{}                                       &      &$\F_{\lambda,\Q}$&                   \\
 \hline
Gumbel dom.  &                \mcmr{4}{|c|}{$\F_0$}                     &$L=\lambda L^*$ &   \mcmr{3}{|c|}{None}               \\
 $L^*(n)$        &                \tmc{4}{|c|}{}                            &  $\F_{\lambda}$   &    \tmc{3}{|c|}{}                   \\
 \hline
Fr\'echet dom. &\mcmr{3}{|c|}{$\F_0$}                  &\tmr{$\F_{\lambda\zeta}$}&          \mcmr{4}{|c|}{None}                      \\
 $\zeta>0$    &\multicolumn{3}{|c|}{}                 &                         &          \multicolumn{4}{|c|}{}                   \\
 \hline
Log.~powers &\tmr{~$\F_\alpha$~} & $\Q=-\alpha,\, L=\lambda L^*$ &                           \mcmr{6}{|c|}{None}                                \\
$\alpha>0, L^*(n)$         &               & $F_{\lambda,-|\Q|}$ &                       \multicolumn{6}{|c|}{}                              \\
\hline
\end{tabular}
}
\caption{Classification of the limit distributions according to the functional
dependence of the exponent $q_n$. The function $L_d(n)$ is a slowly varying
function that goes to infinity with $n$. The slowly varying function $L^*(n)$
characterizes the parent law --see Eq.~(\ref{def-Lstar}).
'None' means that no (non-degenerate) limit distribution emerges.
}
\label{tab:power_ordering}
\end{table}

\begin{table}
\center{
\begin{tabular}{|c|c|c|c|}
\hline
Parent law       & \multicolumn{3}{c|}{Limit of the ratio $q_n/q_n^*$}  \\ \cline{2-4}
                 & $q_n/q_n^* \rightarrow 0$  & $q_n/q_n^* \rightarrow \llambda >0$ & $q_n/q_n^* \! \rightarrow \!+\!\infty$ \\  \hline  
Fr\'echet domain, $\zeta>0$      & Gumbel $\F_{0}$       & Fr\'echet $\F_{\llambda}$              & None \\ \hline
Gumbel domain,  $\F_0$         & Gumbel $\F_{0}$       & Fr\'echet $\F_{\llambda}$               & None \\ \hline
Weibull domain, $\zeta<0$      & Weibull $\F_{\zeta}$  & Non-standard $\F_{\llambda,|\zeta|}$ & None \\ \hline
Logarithmic power, $\alpha>0$  & Fr\'echet $\F_{\alpha}$ & Non-standard $\F_{\llambda,-\alpha}$ & None \\ \hline
\end{tabular}
}
\caption{Classification of the limit distributions according to the
limit of the ratio $q_n / q_n^*$ when $n \rightarrow \infty$.}
\label{tab:transitions}
\end{table}

In conclusion, our heuristic analysis brought to light the existence of transitions between the standard and non-standard attraction domains when modifying the $n$-dependence
of the power transformation $q_n$ (or $q(s)$).
As a byproduct, we also obtained that power transformations can be used
to ``renormalize'' laws beyond the standard attraction domains
(a typical example being logarithmic-power laws) in such a way that
they converge to a non-degenerate distribution.
This set of transitions is summarized in Table~\ref{tab:power_ordering}.
A somehow more compact presentation can be obtained by
using the limit of the ratio $q_n/q_n^*$ in order to classify the limit distributions,
as seen in Table~\ref{tab:transitions}.

\section{ Alternative approach to the characterization of non-standard attraction domains}
\label{sec:transition}

\subsection{Rescaling factors for transformed maximum}

In the previous section, heuristic arguments have shown that only very specific transitions are possible between the standard and non-standard attraction domains. In order to verify this result, this section will present an alternative approach to the study of the transformed maximum in which we shall tie back the convergence behaviour of the transformed and the non-transformed variables.

Let us consider a random variable $W$ which belongs to the domain of attraction of the limit law $\mathcal{F}_\zeta$. This means that there exists a renormalization sequence $(\alpha_n, \beta_n)$ such that  
\be
\label{eq:conv:NP}
\forall x, \quad F_W(\alpha_n x +\beta_n)^n \conv {n} {+\infty} \mathcal{F}_\zeta(x).
\ee
These renormalization sequences are well-known and expressions are available in the literature \cite{EmbrechtETAL:1997,Galambos:1978}.

Similarly, the $\omega$-transformed variable $M^U_n$ converges in distribution
if and only if there exist $\F$ and a sequence $(a_n,b_n)$ such that 
\be
\label{eq:conv:P}
\forall x, \quad F_W\big(\omega_n(a_n x +b_n)\big)^n \conv {n} {+\infty} \tilde{•}ilde{\F}(x).
\ee 

However, in this extension of the previous problem, no general conditions of convergence are known and the choice of the renormalization sequence becomes more difficult. We propose in the next section to exploit the striking similarity between the two previous equations to obtain convergence conditions for the power transformation in specific cases. 

First, one can remark that the only difference between the two previous equations lies in the term $ (\alpha_n x +\beta_n)$ in \Eqref{eq:conv:NP} which becomes $\omega_n(a_n x +b_n)$ in \Eqref{eq:conv:P}. This similarity suggests a simple way to obtain the convergence in distribution of the transformed maximum by exploiting our knowledge of the renormalization factor of the original distribution. If there exists a renormalization sequence $(a_n,b_n)$ such that
\begin{equation}
\label{cond:seq:norm}
\forall x, \quad \omega_n (a_n x +b_n)\simInf{n} \alpha_n \nu(x) + \beta_n,
\end{equation}
and further assuming that 
\begin{equation}
\label{cond:lim:uniform}
\lim_{n\rightarrow +\infty} F_W(\omega(a_n x + b_n))^n=\lim_{n\rightarrow +\infty} F_W(\alpha_n \nu(x) + \beta_n)^n,
\end{equation}
one obtains
\begin{equation} 
\label{eq:conv:F:qn}
\tilde{\F}(x) = \F_\zeta(\nu(x)).
\end{equation}
Therefore, if the conditions (\ref{cond:lim:uniform}) and (\ref{cond:seq:norm}) are satisfied it is possible to link the transformed limit distribution and the standard limit distribution.

Condition~(\ref{cond:lim:uniform}) corresponds to a quite technical convergence problem. For now, we assume that this condition is satisfied.
We show in \ref{app:convergence} that this condition holds
for our proposed choice of $(a_n,b_n)$.
Condition~(\ref{cond:seq:norm}) is more interesting and through $\nu(x)$ defines the kind of transition. For a power transformation, it can be read as 

\begin{equation}
\label{eq:cond:seq:norm:Q}
\forall x,\quad (a_n x +b_n)^{1/q_n} \simD{n}{+\infty} \alpha_n \nu(x) + \beta_n .
\end{equation}

Considering Table~\ref{tab:transitions}, four distinct transitions should be possible.
For laws belonging to the Gumbel or Weibull domain, one should have either $\nu(x)=x$
for $\lim_{n \rightarrow \infty} q_n/q^*_n=0$ or $\nu(x)= \ln (1+\llambda x)/\llambda $ for
$\lim_{n \rightarrow \infty} q_n/q^*_n=\llambda>0$.
Similarly, in the Fr\'echet domain with parameter $\zeta$,
transitions should appear for exponents $q_n$ converging to a finite value,
considering $\nu(x)= [(1+\llambda x)^{\zeta/\llambda} -1]/\zeta$
(with $\llambda= \lim_{n \rightarrow \infty} q_n / \zeta$), and
for vanishing exponents $q_n$ considering $\nu(x)= [\exp(\zeta x)-1]/\zeta $ .

So, using the insight gained from section~\ref{sec:Forcing}, it is natural to study separately the behaviour of $(a_n x +b_n)^{1/q_n}$ for diverging, vanishing and converging power $q_n$.

\subsection{Diverging powers $q_n$}

In the case where $\lim_{n \rightarrow \infty} q_n=+\infty$, the expression $(a_n x +b_n)^{1/q_n}$ has two distinct asymptotic behaviours which lead to two different convergence regimes.

 First, if $a_n/b_n \conv{n}{+\infty} 0$, one has the asymptotic behaviour
\begin{equation}
(a_n x +b_n)^{1/q_n} \simD{n}{+\infty} b_n^{1/q_n}\left(1+ \frac{a_n}{b_n q_n} x \right).
\end{equation}
The condition \Condref{eq:cond:seq:norm:Q} is then satisfied if
\begin{equation}
\label{eq:cond:NAlt}
\left\{ 
\begin{aligned} 
b_n= \beta_n^{q_n} \\
a_n = b_n q_n \frac{\alpha_n} {\beta_n} \\
\end{aligned}
\right. 
\end{equation}
This choice of $(a_n,b_n)$ is compatible with the assumption $a_n/b_n \conv{n}{+\infty} 0$ if
\begin{equation}
q_n \frac{\alpha_n}{\beta_n} \conv{n}{+\infty} 0.
\end{equation}
The term $\alpha_n/\beta_n$ corresponds to the parameter $\lambda_0(n)$ defined
in section \ref{sec:Forcing}. So using the definition of the characteristic exponent
$q^*_n=1/\lambda_0(n)$ introduced in \Eqref{eq:forcing:qstar}, the previous result states that if $q_n$ is negligible with respect to $q^*_n$ (that is, $\lim_{n \rightarrow \infty} q_n/q^*_n=0$) then the transformed maximum converges towards the same limit distribution as the original maximum. This is the expected result from section \ref{sec:Forcing}.

The second asymptotic behaviour arises when $a_n/b_n$ is a constant. By factorizing $b_n$, one obtains

\begin{equation}
(a_n x+b_n )^{1/q_n} \simInf{n} b_n^{1/q_n}\left(1+ \frac{\ln (1+\frac{a_n}{b_n}x)}{q_n} \right).
\end{equation}
\Condref{eq:cond:seq:norm:Q} is satisfied if:
 
 \begin{equation}
 \label{eq:cond:Alt}
\left\{ 
 \begin{aligned} 
b_n= (\beta_n)^{q_n} \\
a_n =  \lambda b_n \\
\frac{q_n} {q^*_n}  \conv{n}{+\infty} \lambda \\
\end{aligned}
\right. 
\end{equation}
In this case, $\nu(x)= (1+\lambda \ln x)/\lambda$ implies that $X_n$ converges towards the non-standard limit laws. More precisely if $q_n$ is asymptotically equivalent to $ \lambda q_n^*$ then the maximum converges in distribution towards $\F_{\lambda,-\zeta}$. Once again, we recover the results of section \ref{sec:Forcing} for the Weibull and Gumbel domain.

\subsection{Vanishing powers $q_n$}

The next interesting transition appears for vanishing moments.
From Table~\ref{tab:transitions}, we know that an exponential term should appears in $(a_n x + b_n)^{1/q_n}$. The easiest way to obtain this term is to assume that $a_n= b_n q_n \zeta$. Then we have
\begin{align*}
(a_n x+b_n )^{1/q_n}&= (b_n)^{\frac{1}{q_n}} (1+ q_n \zeta x) ^ {\frac{1}{q_n}}\\
	& \simInf{n} (b_n)^{\frac{1}{q_n}} e^{\zeta x}.
\end{align*}
Substituting $\nu(x)= [\exp(\zeta x)-1]/\zeta$, one may also satisfy \Condref{eq:cond:seq:norm:Q} by assuming 
 \begin{equation}
 \label{eq:cond:Vanish}
\left\{ 
 \begin{aligned} 
b_n= \left(\beta_n \right) ^{q_n} \\
\frac{\alpha_n}{\beta_n} \equiv q^*_n  \conv{n}{+\infty} \frac{1}{\zeta} \\
\end{aligned}
\right. 
\end{equation}
 
 It is therefore possible to go from the Fr\'echet domain to the Gumbel domain using any decreasing power transformation, as expected from our heuristic analysis in the previous section. This confirms that the Fr\'echet domains are very unstable under power transformation. Any vanishing power is enough to change a distribution belonging to the Fr\'echet domain to converge towards the Gumbel distribution.

\subsection{Converging powers $q_n$}

In the case of converging powers, one expects to observe only transitions
between Fr\'echet domains with distinct parameters $\zeta$. Considering the possible translation and dilation, one can define without loss of generality $a_n=\lambda b_n$ and $b_n=\beta_n^{q_n}$. Then, one obtains
\begin{equation}
(a_n x+b_n )^{1/q_n}=  \beta_n (1+\lambda x)^{1/q_n},
\end{equation}
and choosing $\nu(x)= [(1+\lambda x)^{\lambda/\zeta} -1]/\zeta$ leads to the condition
\begin{equation}
\label{eq:cond:Converging}
\left\{
\begin{aligned}
 q_n^* \conv{n}{+\infty} \frac{1}{\zeta} \\
\frac{q_n}{q_n^*} \conv{n}{+\infty} \lambda
\end{aligned} \right.
\end{equation}
We observe as expected a transition between the Fr\'echet domain of parameter $\zeta$ and the Fr\'echet domain of parameter $\lambda$. If the choice of $\nu(x)$ can appears to be quite arbitrary, it should be noted that it is merely a consequence of the representation chosen for the Fr\'echet limit laws. A different choice of representation ($F(x)=\exp(- x^{-1/\zeta})$) leads to the far simpler $\nu(x)=x^{\lambda/\zeta}$. However, it is compelling that our renormalization methods have allowed us to shed light on this transition even in this convoluted settings. 

With this transition between Fr\'echet domains, we have recovered all the possible transitions from standard attraction domains to other domains described in Table~\ref{tab:transitions} using only the insight obtained from our analysis of the partial equation of the renormalization flow and  elementary arguments on the renormalization coefficients. As shown in \ref{app:convergence}, these arguments lead to a rigorous proof of the convergence of the transformed maximum.

\section{Conclusion}

In this contribution, the renormalization approach of the problem of maximum
developed in \cite{GyorgyiETAL:2008,GyorgyiETAL:2010,Bertin:2010} has been extended to the case where the underlying variables $W_i$ are subjected to a transformation
$\omega_n$, which depends on the sample size. The reduction of the problem of maximum to a partial differential equation turns out to be a rather straightforward generalization of the standard case and leads in the case of the power transformation
$U_{i,n}=W_i^{q_n}$, to a quite short categorization of the limit distributions.  

Using this categorization, non-standard max-stable laws mirroring the standard limit laws have been brought to light. These new limit laws are closely related to the
standard ones. However the behaviour of the partial differential equation describing
the evolution of the distribution of the maximum is more complex and involves some intriguing interactions between the rate of growth of the power transformation and the tail of the distribution. 

These interactions received further investigations by studying the asymptotic behaviour of the forcing parameters appearing in the partial differential equation of the flow. Using a heuristic argument, it was possible to recover a slight approximation of the standard attraction domain. Moreover, the same argument leads to an interesting description of the attraction domain of a non-standard law, illustrating the existence of specific transitions between the classical limit laws and their mirrors laws, when varying the functional dependence of the power $q_n$. These transitions are associated with a characteristic power scale $q_n^*$. If $ q_n / q_n^* \rightarrow 0$, the power transformation $q_n$ is too slow to influence the convergence of the maximum towards the standard limit distributions. In contrast, if $q_n \sim \lambda q_n^*$ the distribution converges towards a non-standard limit distribution. 
Using insights gained from the partial differential equation of flow, it was then possible to confirm the existence of these transitions, and to investigate their mechanisms using a more direct approach based on a study of the rescaling factors.

\subsection*{Acknowledgements}

Interesting discussions with G.~Gy\"orgyi are hereby gratefully acknowledged.

\appendix

\section{Convergence problems}
\label{app:convergence}

\newcommand{\approxN}{\underset{n\rightarrow+\infty}{\approx}}

The convergence results obtained in section~\ref{sec:transition} are dependent on the condition given in \Eqref{cond:lim:uniform}. This condition is equivalent to  $F((a_n x +b_n)^{1/q_n})^n - F(\alpha_n \nu(x) +\beta_n)^n \rightarrow 0$. This is immediately true if $g_W$ is uniformly continuous. However, this condition is superfluous. Let us define
\begin{align} x_n = \alpha_n \nu(x) +\beta_n,\\
\epsilon_n = \frac{(a_n x +b_n)^{1/q_n}- x_n}{\alpha_n}, \end{align}
It is then possible to show that for the four cases presented in section~\ref{sec:transition}, $\lim_{n \rightarrow \infty} \epsilon_n=0.$
For diverging $q_n$ and $ q_n/q^*_n \rightarrow 0$, \Eqref{eq:cond:NAlt} leads to 
\begin{equation} \epsilon_n  \simInf{n} \left(\frac{1}{q_n}-1 \right) \left( \frac{q_n}{q^*_n} \right), \qquad n \rightarrow \infty.  \end{equation}
In the case  $ q_n \sim \lambda q^*_n$, we have   
\begin{equation} \epsilon_n = \ln(1+\lambda x) \left( \frac{1}{\lambda}- \frac{q_n}{q_n^*}\right) + \mathcal{O}\left(\frac{q_n}{(q_n^*)^2}\right). \end{equation}
Similarly for vanishing $q_n$, \Eqref{eq:cond:Vanish} yields
\begin{equation} \epsilon_n = \left(q_n^*-\frac{1}{\zeta}\right) (e^{\zeta x}-1) + \mathcal{O}(q_n). \end{equation}
Since $q_n\rightarrow 0$ in this case and $q_n^* \rightarrow 1/\zeta$, $\epsilon_n$ is therefore a vanishing quantity.
And lastly, for converging moment and $q_n \sim \lambda q_n^*$ 
\begin{equation} \epsilon_n \approxN \frac{1}{\zeta} \big(( 1+\lambda x)^{1/q_n}-( 1+\lambda x)^{\zeta/\lambda} \big) +\frac{1}{\zeta} - q_n^* \end{equation}
Using $q_n^* \rightarrow 1/\zeta$, this confirms that $\lim_{n\rightarrow \infty} \epsilon_n=0$. 

Moreover, by construction $\lim_{n \rightarrow \infty} n \CF(x_n) \in \R$. It is possible to show that $F((a_n x +b_n)^{1/q_n})^n - F(\alpha_n \nu(x) +\beta_n)^n \rightarrow 0$ is equivalent to 
\begin{equation}
\label{eq:cond:cF}
\lim_{n\rightarrow \infty} \frac{\CF(x_n + \alpha_n \epsilon_n)}{\CF(x_n)} = 1.
\end{equation}

If \Condref{eq:cond:cF} is satisfied, then the convergence in distribution of $M_n^U$
is ensured. In sections~\ref{Proof:Gumbel}, \ref{Proof:Weibull} and \ref{Proof:Frechet}, we verify that this condition holds for all the standard domains. Consequently, the transition described earlier is always valid.

\subsection{Gumbel domain}
\label{Proof:Gumbel}

The Gumbel attraction domain is the harder to characterize. In order to prove the convergence in the general settings, we will temporarily use the standard renormalisation factors
\begin{gather}
\beta_n = \Inv {\CF}(1/n),\\
\alpha_n= E(\beta_n),
\end{gather}
where the function $E(x)$, defined as
$E(x)= \frac{1}{1-F(x)}\int_x^{+\infty} 1-F(t) dt$, satisfies:
\begin{equation}
\label{eq:cond:Gumbel:F}
\begin{gathered}
\lim_{x\rightarrow+\infty} E'(x)=0 \\
\forall r>0, \quad \lim_{x\rightarrow+\infty} \frac {1-F(x+r E(x))} {1-F(x)} = e^{-r}.
\end{gathered}
\end{equation}
One useful property of $E$ is that for any positive real $r$
\begin{equation}
\label{eq:cond:Gumbel:E}
\frac {E(x+r E(x))} {E(x)} \conv{x}{+\infty} 1.
\end{equation} 
Combining Eqs.~(\ref{eq:cond:Gumbel:F}), (\ref{eq:cond:Gumbel:E}) and (\ref{eq:cond:Gumbel:E}) leads to 
\begin{equation}
\begin{aligned}
E(x_n) &= E( \alpha_n x + \beta_n) \\
&= E( \beta_n + x E(\beta_n)) \sim E(\beta_n),
\end{aligned}
\end{equation}
so that $E(x_n) \sim \alpha_n$.
Consequently, $\alpha_n \epsilon_n/E(x_n) \rightarrow 0$, and using this result with \Eqref{eq:cond:Gumbel:F} yields 
\begin{equation}
\begin{aligned}
\CF(x_n + \alpha_n \epsilon_n)&= \CF(x_n+ \frac{\alpha_n \epsilon_n}{x_n E(x_n)} E(x_n) x_n)\\
&\sim e^{-\frac{\alpha_n \epsilon_n}{x_n E(x_n)}}  \CF(x_n)\\
 & \sim \CF(x_n) 
\end{aligned}
\end{equation}     
So \Condref{eq:cond:cF} holds for the Gumbel attraction domain.

\subsection{Weibull domain}
\label{Proof:Weibull}

As stated in \Eqref{eq:domain:Weibull}, we have for the Weibull domain
 \begin{equation}
 \CF(x)= L\big((x_F-x)^{-1}\big)\, (x_F-x)^{-1/\zeta}.
\end{equation}
Moreover, using the definition of the renormalisation factor leads to 
\begin{equation}
\frac{x_F-\beta_n}{\alpha_n} \simInf{n} \frac{(x_F-\beta_n) \CF(\beta_n)}{\CF'(\beta_n)}.
\end{equation}
Hence, with the assumption that $ L'/L \rightarrow 0$ we have
\begin{equation} \lim_{n \rightarrow \infty} \frac{x_F-\beta_n}{\alpha_n} = \zeta. \end{equation}
Therefore, it is possible to show that 
\begin{equation} x_F-( x_n + \alpha_n \epsilon_n) \sim \alpha_n x \left(1+\frac{\zeta}{x}\right).
\end{equation}
The properties of slowly varying functions yield, for $n \rightarrow \infty$
\begin{equation}
\begin{aligned}
\CF(x_n + \alpha_n \epsilon_n)&= L\left(\frac{1}{x_F-x_n- \alpha_n \epsilon_n}\right)(x_F-x_n-\alpha_n \epsilon_n)^\zeta\\
& \sim L\left(\frac{1}{\alpha_n x} \right) (\alpha_n x(1+\zeta/x))^{-\frac{1}{\zeta}} \\
&\sim  \CF(x_n)
\end{aligned}
\end{equation}
Hence \Condref{eq:cond:cF} holds for the Weibull domain.

\subsection{Fr\'echet domain}
\label{Proof:Frechet}

Within the Fr\'echet domain, the proof is immediate using $\alpha_n \epsilon_n/x_n \rightarrow 0$ which is directly implied by $\lim_{n \rightarrow \infty} \epsilon=0$ and $\lim_{n \rightarrow \infty} q_n^* >0$. We have from \Eqref{eq:domain:Frechet}
\begin{equation}
\begin{aligned}
\CF(x_n+\alpha_n \epsilon_n)=(x_n+\alpha_n \epsilon_n)^{-1/\zeta} L(x_n+\alpha_n \epsilon_n) \\
\sim x_n^{-1/\zeta} L(x_n) =\CF(x_n).
\end{aligned}
\end{equation} 
As a result, \Condref{eq:cond:cF} is also satisfied for the Fr\'echet class.

\section*{References}
\bibliography{biblio_evs}

\end{document}